\documentclass[sigconf]{acmart}

\AtBeginDocument{%
  \providecommand\BibTeX{{%
    \normalfont B\kern-0.5em{\scshape i\kern-0.25em b}\kern-0.8em\TeX}}}

\copyrightyear{2024}
\acmYear{2024}
\setcopyright{rightsretained}
\acmConference[ICSE-SEIS'24]{Software Engineering in Society}{April 14--20, 2024}{Lisbon, Portugal}
\acmBooktitle{Software Engineering in Society (ICSE-SEIS'24), April 14--20, 2024, Lisbon, Portugal}\acmDOI{10.1145/3639475.3640112}
\acmISBN{979-8-4007-0499-4/24/04}

\usepackage{svg}
\usepackage{amsmath}
\usepackage{color}
\usepackage{url}
\usepackage{colortbl}
\usepackage{booktabs}
\usepackage{hhline}
\usepackage{multirow}
\usepackage{enumerate}
\usepackage{xspace}
\usepackage{hyperref}
\usepackage{cleveref}
\usepackage{quoting}
\usepackage{graphicx}

\usepackage[nolist]{acronym}

\definecolor{lightgray}{rgb}{0.83, 0.83, 0.83}
\definecolor{darkgray}{rgb}{0.66, 0.66, 0.66}
\definecolor{lightergray}{rgb}{0.945,0.945,0.945}

\newcommand{\quotes}[1]{`#1'}

\begin{document}

\title{An Empirical Study on Compliance with Ranking Transparency in the Software Documentation of EU Online Platforms}

\author{Francesco Sovrano}
\authornote{Both authors contributed equally to the paper}
\affiliation{%
	\institution{University of Zurich}
	\city{Zurich}
	\country{Switzerland}
}
\email{francesco.sovrano@uzh.ch}

\author{Michaël Lognoul}
\authornotemark[1]
\affiliation{%
	\institution{University of Namur (CRIDS, NADI)}
	\city{Namur}
	\country{Belgium}
}
\email{michael.lognoul@unamur.be}

\author{Alberto Bacchelli}
\affiliation{%
	\institution{University of Zurich}
	\city{Zurich}
	\country{Switzerland}
}
\email{bacchelli@ifi.uzh.ch}

\begin{abstract}
Compliance with the European Union's Platform-to-Business (P2B) Regulation helps fostering a fair, ethical and secure online environment. However, it is challenging for online platforms, and assessing their compliance can be difficult for public authorities. This is partly due to the lack of automated tools for assessing the information (e.g., software documentation) platforms provide 
concerning ranking transparency. 
Our study tackles this issue in two ways. First, we empirically evaluate the compliance of six major platforms (Amazon, Bing, Booking, Google, Tripadvisor, and Yahoo), revealing substantial differences in their documentation. Second, we introduce and test automated compliance assessment tools based on ChatGPT and information retrieval technology. These tools are evaluated against human judgments, showing promising results as reliable proxies for compliance assessments. Our findings could help enhance regulatory compliance and align with the United Nations Sustainable Development Goal 10.3, which seeks to reduce inequality, including business disparities, on these platforms.\\
\textbf{Data and materials:} \url{https://doi.org/10.5281/zenodo.10478546}.
\end{abstract}

\ccsdesc[500]{Software and its engineering~Documentation}
\ccsdesc[500]{Information systems~Data analytics}
\ccsdesc[500]{Applied computing~Law}
\ccsdesc[300]{Computing methodologies~Natural language processing}

\keywords{
    Software Documentation,
    EU Regulations,
    Compliance Assessment,
    Ranking Transparency,
    Explainability,
    Online platforms
}

\maketitle

\begin{acronym}
    \acro{AI}{Artificial Intelligence}
    \acro{P2B}{Platform-to-Business}
    \acro{DoX}{Degree of Explainability}
\end{acronym}

\section*{Lay Abstract}


In the digital world, software powers online platforms such as marketplaces and search engines. For instance, software determines the display order of products on Amazon or hotels on Booking.com. Understanding these systems is vital for businesses, yet the explanations can be complex.
The European Union mandates online platforms to reveal the main parameters of their ranking algorithms, aiming for fairness and reducing inequalities. However, ensuring these platforms comply is difficult due to the lack of a standard evaluation method.
Our study examined explanations from Google, Yahoo, Bing, Amazon, Booking, and Tripadvisor. We gathered insights from three legal experts and over a hundred people to gauge the clarity of these explanations, finding significant variations in clarity across platforms.
Reviewing these explanations manually is cumbersome. To streamline this, we developed automated tools and compared their effectiveness against human assessments. Our discussion revolves around how these tools can enhance transparency on online platforms.


\section{Introduction}

In today's digital landscape software is more than just lines of code, it is a driving force behind economic activities and business operations \cite{DBLP:conf/amcis/GajardoP19,DBLP:conf/icis/BrynjolfssonO12,brynjolfsson2002understanding}. This impact is particularly evident in the online platforms' ecosystem, especially regarding intermediation services (e.g., marketplaces) and search engines. The software used by the providers of these types of services plays a significant role in shaping downstream business dynamics~\cite{busch2023algorithmic}. As software increasingly becomes the backbone of society, attention is shifting towards its interaction with people. One critical element of this relationship is the need for clear and transparent software documentation, often not provided by default to those who interact with software.

To address this issue, the European Union has introduced legislation such as the \ac{P2B} Regulation (EU) 2019/1150~\cite{eurlex2019}. Among other requirements, this Regulation imposes on platforms to provide easily accessible, comprehensible, and detailed documentation explaining how ranking mechanisms work and affect sellers, including business and corporate website users (Article 5 and Recital 25~\cite{EU2023P2B}). Such a measure aims to create a more equitable environment for businesses that rely on online platforms to connect with consumers. This resonates with Target 10.3 of the United Nations Sustainable Development Goals~\cite{UN2023}, which aims to ensure equal opportunity and reduce inequalities within and among countries.

However, producing documentation adhering to the \ac{P2B} Regulation is challenging for online platforms, as recently pointed out by the European Commission (EC)~\cite{EU2023P2B}. 
Additionally, monitoring the quality of documentation is challenging for platforms and regulators, affecting compliance assessment, law effectiveness, and social well-being.
This is largely due to the lack of standardized, automated tools for assessing compliance 
across different platform providers. Hence, given the absence of standardized assessment tools, we hypothesize to encounter varying levels of compliance among platforms regarding ranking transparency.

To explore the extent to which online platforms actually produce adequate documentation, as intended by the law, we first conducted a manual assessment involving three legal experts from Belgium (with respectively 7, 6 and 5 years of experience in legal research) and more than 130 participants. Our focus is on evaluating whether the software documentation from key online intermediation services (i.e., Amazon, Tripadvisor, and Booking) and online search engines (i.e., Google, Bing, and Yahoo) genuinely aids businesses in understanding ranking mechanisms and can fulfill the objective pursued by the Regulation.
We expect this empirical study to validate our initial hypothesis, affirming the variance in the quality of explanations provided by different platforms.

As a second step, we investigate whether automated tools can effectively assess compliance as a proxy for traditional surveys. Given the recent popularity of generative AI, we experiment with ChatGPT as a baseline assessment tool. We also propose an alternative approach, which combines ChatGPT with answer retrieval technologies, offering a more transparent assessment based on a novel metric, DoX~\cite{SOVRANO2023110866,sovrano2022how}, designed to quantify the degree of explainability of documentation. 
%
We test these automated tools to evaluate their correlation with human judgments, as obtained from our empirical study. We aim to identify these systems' strengths and weaknesses in predicting compliance and explainability.



The primary contributions of this work include:
\begin{enumerate}
    \item A checklist derived from the \ac{P2B} Regulation and European Commission guidelines for platform documentation quality assessment.
    \item A method to automate assessment using the aforementioned checklist through ChatGPT.
    \item A more deterministic, non-hallucinatory, and transparent assessment tool than ChatGPT, which is based on a theory of explanations from Ordinary Language Philosophy~\cite{achinstein2010evidence,sovrano2022explanatory} and answer retrieval technology.
    \item An in-depth analysis of ranking documentation from six major online platforms involving three legal experts and over 130 potential customers. 
    \item Methods that closely align with expert views on compliance, illustrating their potential for ongoing monitoring by authorities and the public.
    \item Source code for the automated tools and referenced data, including manual assessments and checklist~\cite{replication_package}.
\end{enumerate}

Considering the long-term societal ramifications, we argue that effective automated tools can lead to the development of more compliant, transparent software documentation and, consequently, more equitable online platform ecosystems for businesses. This potential transformation could significantly impact the pursuit of the UN Sustainable Development Goal 10.3, which focuses on reducing inequality.


\section{Related Work} \label{sec:related_work}

This section compares our analysis of the \ac{P2B} Regulation and the proposed assessment tools with existing literature.

In 2022, the Observatory on the Online Platform Economy \cite{EUObservatory2023} conducted a study for the European Commission, focusing on the \ac{P2B} Regulation \cite{PlatformObservatory2021}. It examined terms and conditions of platforms and analyzed feedback from business users. Despite changes by major platforms like Amazon, Facebook, and Google, over half of the business users did not see improvements in the transparency of terms and conditions and ranking practices.

Subsequently (September 2023), the European Commission published a review on the implementation of Regulation (EU) 2019/1150 \cite{EU2023P2B}. Their methodology included interviews and analysis of online platform terms, finding that compliance with the \ac{P2B} Regulation remains a challenge despite improvements in transparency. In particular, the European Commission noted that \quotes{in many cases, the basic information provided, for example, on ranking, is potentially insufficient} (p.9).

In contrast to the approach we propose in this paper, the review of both the European Commission~\cite{EU2023P2B} and the Observatory~\cite{PlatformObservatory2021} employed manual assessments. They do not incorporate a detailed quantitative analysis concerning the \quotes{quality of the explanations} found in the terms and conditions.

Additionally, \citeauthor{lukovic2021information} \cite{lukovic2021information} examined the context and actors of platform rankings. This study details the ranking transparency obligations set by the \ac{P2B} Regulation and offers critical insights into the framework's limitations. \citet{bibal2021legal} situate the \ac{P2B} Regulation's ranking parameters within a broader EU context that mandates explanations for AI systems. This work sheds light on the intentions of EU legislators and specifies the type and extent of information essential for technical compliance. The approaches of both \citet{lukovic2021information} and \citet{bibal2021legal} differ from ours as they neither evaluate platforms' adherence to the Regulation nor introduce tools to enhance regulatory compliance. These elements are actually absent from existing literature, to our knowledge.


In our research, we focus on the documentation associated with the \ac{P2B} regulation. In particular, we evaluate the potential advantages and limitations of leveraging ChatGPT's prompt engineering for assessment purposes. While ChatGPT has previously been harnessed for various tasks such as evaluating student responses~\cite{DBLP:journals/corr/abs-2305-12962} or human behavior~\cite{DBLP:journals/corr/abs-2303-01248}, we instead apply it to assessing software documentation. 

Distinguishing our work from prior studies, we propose a method designed to evaluate also the clarity and explainability of software documentation, an aspect often overlooked. This undertaking complements the growing academic interest in developing both semi-automatic and fully automatic techniques to uphold software documentation quality~\cite{DBLP:conf/ease/GarousiGMRS13, DBLP:journals/infsof/GarousiGRZMS15}.

\section{Background} \label{sec:background}

This section explores the EU's \ac{P2B} Regulation, fostering fairness in online platform-business interactions, alongside the European Commission's ranking transparency guidelines. It also introduces the DoX metric for measuring explainability, setting the stage for later discussions on assessment tools.

\subsection{P2B Regulation \& Guidelines} \label{sec:background:p2b_regulation}

In 2019, the EU adopted the \ac{P2B} Regulation~\cite{eurlex2019} to improve fairness and transparency for online platforms in their dealings with businesses. As noted by \citet{prastitou2021notion}, it targets \textit{online intermediation services}, i.e., services that allow business users to offer products to consumers and are based on contractual relationships (see Art. 2.2). The Regulation also covers \textit{online search engines}, where users can search the internet (Art. 2.5).

The Regulation identifies two primary beneficiaries: business users and corporate website users of platform services. \textit{Business users} are private entities acting commercially, offering goods or services to consumers via online intermediaries (Art. 2.1); \textit{Corporate website users} are entities using an online interface, like a website or software, for professional activities to offer goods or services to consumers (Art. 2.7).

The Regulation requires online intermediation service providers to keep their terms and conditions easily accessible to business users, written in clear language (Art. 3.1). They must disclose certain information therein, such as the main criteria for ranking offers to consumers, and the reasons for their importance (Art. 5.1), as mentioned by \citet{hacker2022regulating}.
Similarly, online search engines must provide a clear, public description of the main parameters that determine ranking and their significance (Art. 5.2 \cite{EU2023P2B}).



The EU legislator seeks to ensure online platforms offer explanations for business and corporate website users about ranking mechanisms \cite{EU2023P2B,EUObservatory2023}. This includes understanding how rankings consider offer characteristics, relevance to consumers, and for search engines, website design traits (Art. 5.5). Explanations must also clarify the ability to influence ranking through payments and their impacts (Art. 5.3).

The primary objectives of the legislator in enforcing ranking transparency are twofold: \textit{(1)} to empower businesses in the presentation of their offers online, enhancing competition in downstream markets (Recital 24) and \textit{(2)} to enable businesses to compare ranking practices, promoting competition among online platforms in upstream markets (Recital 24). These objectives align with the United Nations Sustainable Development Goal 10.3 on reducing inequality.

To enhance compliance with ranking transparency, the Regulation required the European Commission to adopt Guidelines (Art. 5.7). In short, the European Commission's Guidelines~\cite{EC2020P2BGuidelines} are a non-binding text (pt. 9~\cite{EC2020P2BGuidelines}) that elaborates on the content of the binding \ac{P2B} Regulation.

The European Commission's Guidelines emphasize that online platforms' explanations should be tailored to the needs and technical abilities of average users, differing based on the service type (pt. 17). The descriptions should offer more than a mere list of main parameters and give a secondary layer of information (pt. 22). Descriptions should not be overly brief or misleading (pt. 98). Moreover, the explanations should cover all main ranking parameters, indicating exhaustive lists (pt. 24).


Regarding the form that the explanations should take, both the Regulation and the Guidelines use the term 'drafted' in English, which refers to written content.\footnote{For definitions and examples of the transitive verb form of "draft," refer to the Cambridge Dictionary at \url{https://dictionary.cambridge.org/fr/dictionnaire/anglais/draft}.} To verify this, we looked into the French versions of both texts. The French translation of the Regulation refers to redacted explanations (`rédigées') within their terms and conditions for online intermediation services (Art. 3.1 and 5.1). For search engines, the exact text uses the words `to state' (`énoncée'), which do not imply the use of one form over another (Art. 5.2). The Guidelines, however, refer at some point to the `redaction' of the explanations for both types of providers, potentially implying the use of a written medium in all cases (i.e., ``lorsqu'ils rédigent la description [...]", pt. 99). We also looked into the Dutch version of the Regulation, which refers to `redacted' content (`opgesteld') regardless of the type of service provider (Art. 3.1 and 5.2).


\subsection{How to Measure Explainability} \label{sec:background:dox}

The Degree of eXplainability, abbreviated as DoX, is a model-agnostic metric introduced by \cite{SOVRANO2023110866,sovrano2022how} to measure the clarity and depth of explanations within a text. Grounded in Achinstein's theory \cite{achinstein2010evidence}, DoX equates the act of explaining to the process of answering fundamental questions, referred to as \textit{archetypes}. The more archetypal questions a text can answer, the higher its explainability rating. Notably, DoX can evaluate any content written in natural language once what is to be explained is defined.

In particular, DoX measures similarity, exactness, and fruitfulness, as articulated by Carnap \cite{novaes2017carnapian}. These measurements are then merged to produce a numerical (DoX) score, providing the means for a comparison of the quality of different explanations.

To calculate DoX scores, the \textit{relevance} of a text snippet is assessed and aggregated based on its relation to a collection of archetypal questions informed by linguistic theories (see \cite{SOVRANO2023110866}). Specifically, these relevance measurements are obtained employing pre-trained \textit{deep language models} tailored for general-purpose answer retrieval \cite{karpukhin2020dense,bowman2015large}. For examples on how DoX works, see the paper \cite{SOVRANO2023110866}.

\section{Study \& Checklist Design} \label{sec:case_study_n_checklist}



In this section, we detail our methodology for selecting platform documentation, which we assess later on. We also describe our method and process for developing an assessment checklist.

\subsection{The Study} \label{sec:case_study_n_checklist:data_collection}

We independently investigate, beyond the studies and reports outlined in \Cref{sec:related_work}, the adequacy of documentation provided by several prominent online platforms as mandated by the Regulation. Our analysis focuses on software documentation from three major online intermediation services (Amazon, Tripadvisor, and Booking) and three online search engines (Google, Bing, and Yahoo).

We selected these platforms for their industry representativeness and the audience profile they serve, which is very broad and allowed us to perform a large-scale manual assessment (cf. Section \ref{sec:results:prolific}).

In compliance with the \ac{P2B} Regulation, intermediation services are expected to provide their users with readily accessible information within their terms and conditions (see Art. 3.1 and 5.1), and search engines within a 'publicly available description' (see Art. 5.2). However, our investigation revealed a convoluted landscape. Information on ranking, which should be easily accessible, was often dispersed across multiple pages, thus requiring users to navigate through multiple re-directions. In addition, not all the selected intermediation services consistently included this crucial data in their terms and conditions, adding to the complexity of the retrieval process. 

Most platforms, however, maintained a primary web page detailing their ranking practices. This became our starting point. We confined our search to these pages and any information within one click's distance. 
We also observed the use of audiovisual content by some platforms, in particular by Amazon and Google, to explain ranking mechanisms. However, these videos were excluded from our study, in line with the developments held above on the provision of ranking information in a written form (see \Cref{sec:background:p2b_regulation} and Art. 3.1, 5.1, and 5.2). 

For a quantitative perspective, \Cref{tab:platform_stats} presents a breakdown of the documents we retrieved from the six platforms, detailing the number of associated links and the average word count per document. Our online repository provides further details~\cite{replication_package}.

\begin{table}[htb]
    \centering
    \caption{Statistics on retrieved documents per platform.}
    \label{tab:platform_stats}
    \begin{tabular}{lrr}
        \hline
        Platform & No. of Links & Avg. Words/Doc \\
        \hline
        Amazon & 5 & 434.40 \\
        Bing & 16 & 964.06 \\
        Booking & 7 & 579.42 \\
        Google & 52 & 1679.50 \\
        Tripadvisor & 10 & 1653.90 \\
        Yahoo & 3 & 174.00 \\
        \hline
    \end{tabular}
\end{table}

\subsection{The Checklist} \label{sec:case_study_n_checklist:checklist}

To assess the documentation quality of selected platforms, we created a checklist of essential questions, all needing positive responses for an operator to meet ranking transparency requirements. These questions were crafted by a legal expert and co-author of this study, an academic researcher with expertise in EU law on information and communication technologies and 6 years of experience. This was done using an inductive coding approach \cite{fereday2006demonstrating}, allowing the checklist questions to emerge from the P2B regulation and guidelines by the European Commission. The checklist was developed based on these materials through the following steps.


Firstly, we focused on terms like `ranking' and `main parameter', which are legally defined and central to the \ac{P2B} Regulation's explanations. However, they might be unclear to some users. Therefore, we drafted two questions to verify whether these terms were clarified in the platform's documentation (cf. Q\ref{q1} and Q\ref{q2}).

Secondly, we incorporated each of the individual elements explicitly required by the Regulation within its own question, taking into account the wording of the legal text. These questions, therefore, address matters such as \textit{(i)} the provision of the main parameters used for ranking (cf. Q\ref{q3}), \textit{(ii)} how ranking considers the characteristics of the goods and services offered to consumers (cf. Q\ref{q5}), \textit{(iii)} the extent to which such characteristics are taken into account (cf. Q\ref{q6}), \textit{(iv)} the existence of payments to influence ranking (cf. Q\ref{q9}), and \textit{(v)} the effects of such payments (cf. Q\ref{q10}).

However, the Regulation differentiates between two types of platform operators: online intermediation services and online search engines. Both are covered, but they face different requirements, as in Articles 5.1 and 5.2, and point 7 of the Guidelines. Therefore, we have created a different set of yes-or-no questions for each platform type. Importantly, this task was led by solely paying attention to the explicit requirements contained in the Regulation. 

Thirdly, we drafted additional questions based on the content of the European Commission's guidelines. 
Whenever these guidelines explicitly raised new questions, proposed nuances to the requirements of the legal text, or gave best practices to comply with the \ac{P2B} Regulation and its objectives, we designed new questions to take these elements into account and incorporated them into our checklist. These questions relate to elements like \textit{i)} the explanation of what is most important in determining ranking (cf. Q\ref{q12}) and \textit{ii)} the internal process conducted by the platforms to determine the main parameters (cf. Q\ref{q14}).

Fourthly, we revised certain questions' wording, without altering their meaning, to address issues in automated answer retrieval from platform documentation. For example, Q\ref{q11} initially included \quotes{all the main parameters} (cf. Guidelines, pt. 24). However, this phrasing led to issues in automated scoring, as no platform explicitly states that the provided list of parameters is exhaustive. To address this, we modified Q\ref{q11} by omitting the word 'all', thereby improving the precision of automated assessments.

The list of questions for online intermediation services is: 
\begin{enumerate}[Q1]
    \item Does the documentation explain how 'ranking' is defined/define 'ranking'? \label{q1}
    \item Does the documentation explain how 'main parameter used for ranking' is defined/define 'main parameter used for ranking'? \label{q2}
    \item Does the documentation provide the main parameters used for determining ranking? \label{q3}
    \item Does the documentation explain why certain parameters are considered as the main ones for determining ranking instead of others? \label{q4}
    \item Does the documentation explain how the ranking mechanism considers the characteristics of the goods and services offered to consumers? \label{q5}
    \item Does the documentation explain the extent to which the ranking mechanism considers the characteristics of the goods and services offered to consumers? \label{q6}
    \item Does the documentation explain how the ranking mechanism considers the relevance of the characteristics of the goods and services, for consumers? \label{q7}
    \item Does the documentation explain the extent to which the ranking mechanism considers the relevance of the characteristics of the goods and services, for consumers? \label{q8}
    \item Does the documentation explain the possibilities to influence ranking against direct or indirect payment (if any)? \label{q9}
    \item Does the documentation explain the effects of payments, on ranking (if any)? \label{q10}
    \item Does the documentation explain how the ranking mechanism works and, in particular, what the main parameters used are? \label{q11}
    \item Does the documentation explain what is most important in determining ranking? \label{q12}
    \item Does the documentation explain why specific parameters were selected as the main factors in determining the ranking of goods or services? \label{q13}
    \item Does the documentation explain the internal process conducted by the provider to determine the main parameters for the ranking of goods or services? \label{q14}
    \item Does the documentation explain how users can improve the ranking of their goods or services? \label{q15}
    \item Does the documentation explain how users can alter the ranking of their products or services through direct or indirect payments to the provider, and what effect this has? \label{q16}
    \item Does the documentation explain what the business logic behind allowing users to affect the ranking of their products or services through payments is, and what the potential consequences of this are? \label{q17}
\end{enumerate}

The set of questions for online search engines is mainly the same but differs for three questions. 
Question \ref{q4} has been changed to \quotes{Does the documentation provide the relative importance of the different main parameters used in determining ranking?}. The difference between both questions lies in two elements. Firstly, search engines do not have to provide contrastive explanations, as they are not explicitly required to explain why some main parameters were selected \textit{as opposed to others}. Secondly, search engines have to specify the relative importance of their main parameters, whereas intermediation services do not. 
Moreover, in accordance with the legal text and to address the differences between types of services at stake, we added two new questions: 
\begin{enumerate}[Q1]
    \setcounter{enumi}{17}
    \item Does the documentation explain how the ranking mechanism considers the design characteristics of the websites? \label{q18}
    \item Does the documentation explain the extent to which the ranking mechanism considers the design characteristics of the websites? \label{q19}
\end{enumerate}

Although some of the questions, within each subset, might seem quite similar at first sight, they all call for (at least slightly) different answers, as they contain nuances that originate from the European Commission's Guidelines. For instance, questions number \ref{q4} and \ref{q13} appear to be very similar. However, question \ref{q4} puts emphasis on the contrast between the parameters selected as main ones and other parameters, whereas question \ref{q13} does not. Similarly, questions number \ref{q3} and \ref{q11} seem quite close to each other. Yet, question \ref{q3} (based on the binding text of the Regulation) only refers to the main parameters used, while question \ref{q11} (based on the non-binding guidelines) additionally considers how the ranking mechanism works in general.

\section{Automated Assessment} \label{sec:algorithms}


This section delves into different strategies that use checklists to assess compliance. While conventional methods, which require manual examination of checklists against extensive documents (see Section \ref{sec:case_study_n_checklist}), are not only lengthy but also susceptible to personal interpretations (as shown in Section \ref{sec:results:experts}), automated assessments offer regulators and public authorities significant advantages. Automated systems provide consistency in evaluation, greatly reduce the time required for assessment, and minimize the potential for human error or subjective biases. For enforcement agencies, this means faster, more reliable evaluations, ensuring more effective oversight and public trust.

Hence, we hereby discuss two computational approaches. The first uses off-the-shelf ChatGPT, while the second is based on answer retrieval technology and the DoX metric (see Section \ref{sec:background:dox}) for more robust and transparent evaluations.

\subsection{ChatGPT-based Assessment} \label{sec:algorithms:chatgpt}


Using ChatGPT, we developed an automated algorithm to determine if documentation aligns with the criteria detailed in Section \ref{sec:case_study_n_checklist}. Given that ChatGPT models possess inherent limitations regarding input size, our system design had to account for such constraints. For example, GPT-4 limits inputs to 8192 tokens, while the GPT-3.5-turbo-16k version allows for up to 16385 tokens.\footnote{The GPT-4 and 3.5 versions cited are the 0613 variants, a snapshot taken on June 13th, 2023.} 
Consequently, our approach breaks the software documentation into segments that adhere to GPT's token restrictions. Each section is then individually assessed against the checklist and is given a score between 1 (indicating non-adherence) and 5 (indicating complete compliance), as referenced in \ref{sec:results:experts}. After this assessment, the algorithm consolidates these scores, selecting the highest score for every checklist point.

We used the following prompt to instruct the ChatGPT model:\smallskip

\begin{quoting}[font=itshape]
\noindent Your task is to assess the compliance of this documentation based on the following question. Conduct a compliance assessment, focusing on both the technical and legal requirements. \\
Your assessment should start with a numerical score from 1 to 5, where 1 indicates the question is not answered at all and 5 indicates it's perfectly answered. Following the score, provide a brief explanation highlighting the strengths or weaknesses in addressing the question. Consider the completeness, clarity, and legal implications in your explanation. \\
For example, your assessment might look like: 
'Score: 3. Explanation: The question was only partially answered. While the technical aspects are covered, it lacks legal disclosures.' \\
Question:
\{question\} \\
Documentation:
\{chunk\}
\end{quoting}\smallskip

Additionally, we set the model's 'temperature' to zero to reduce randomness, thus ensuring more deterministic outcomes.

This approach showcases an elementary form of prompt engineering\footnote{Prompt engineering involves formulating precise task descriptions and examples to steer the responses of language models \cite{henrickson_prompting_2023}.}, aiming to direct the model towards expected outcomes. Yet, the foundational prompt has inherent challenges. One major issue is that segmenting extensive documentation can potentially affect the integrity of the evaluation since pertinent data might be split across various segments. The importance of this becomes evident when considering that our system employs a max operator to amalgamate scores, neglecting the differential significance of each segment. Additionally, the simplistic nature of this prompt renders the model prone to producing fabricated outputs \cite{alkaissi2023artificial,azamfirei_large_2023}.

Considering these challenges, we have devised the improved strategy presented in \Cref{sec:algorithms:dox}. This approach incorporates advanced prompt engineering and more transparent answer retrieval mechanisms to effectively summarize lengthy documentation.

\subsection{Assessing Explanation Quality with DoX} \label{sec:algorithms:dox}

A key limitation of the standard ChatGPT-based approach presented in \Cref{sec:algorithms:chatgpt} is its inability to quantitatively assess the quality of explanations.
Capturing this characteristic is quintessential to determine if the documentation provides clear explanations, especially when it comes to the legal requirements of the \ac{P2B} Regulation. For instance, a platform's documentation stating \quotes{Rankings are based on algorithms} might be accurate, but lacks the necessary depth. According to the principles of Ordinary Language Philosophy~\cite{achinstein2010evidence,sovrano2022explanatory}, a proper explanation would provide more context, such as \quotes{Rankings are determined by algorithms that consider factors like user engagement, relevance, and content quality.}

As highlighted in \Cref{sec:background:dox}, the DoX metric is designed to assess such qualities and estimate the degree of explainability of a given documentation. Thus, we have devised a technique that combines a transparent, theory-based answer retrieval system (optimized for parsing lengthy documentation efficiently), ChatGPT (to sift through and paraphrase the retrieved answers), and the DoX metric (to estimate the explanatory power of a text).

In particular, we employ the same neural retrieval model \cite{DBLP:conf/jurix/SovranoPV20} used by the DoX algorithm (cf. Section \ref{sec:background:dox}) to extract the answers to the checklist questions from the documents and ChatGPT to refine them into comprehensive explanations. Then, DoX is used to measure the explanatory quality of the answers to the checklist questions.
Using the earlier example for clarity: while both answers \quotes{Rankings are based on algorithms} and \quotes{Rankings are determined by algorithms that consider factors like user engagement, relevance, and content quality} would achieve similar relevance ratings, the first would score lower on the DoX metric, indicating its limited explanatory depth.

Notably, a checklist is framed for binary responses, but common answer retrieval tools (such as the one we adopted) are more suited for open-ended queries, being designed for paragraph-length responses. As a result, we had to modify the checklist questions from Section \ref{sec:case_study_n_checklist} to accommodate a more open-ended format. 

Converting these questions often means reshaping them to target the specific information needed. For instance, take the transition from the closed-ended question, \quotes{Does the documentation explain how 'ranking' is defined?} to its open-ended counterpart, \quotes{How is 'ranking' defined?}. While both queries focus on the interpretation of 'ranking', the second one elicits a more comprehensive answer rather than a mere yes or no.

Here is a more detailed breakdown of the steps involved in this conversion process:
\begin{enumerate}
    \item \textit{Identify the Key Element}: First, identify the essential piece of information that the original question aims to find out. This is often buried in clauses like \quotes{Does the documentation include...}.
    \item \textit{Re-frame as Direct Question}: Next, rephrase the question to ask directly about that key element.
\end{enumerate}
With these modifications, the questions become more straightforward and easier for the information retrieval system to address.

Subsequently, the answer retriever selects the top 20 best answers from documentation paragraphs. Each answer gets a pertinence score from 0 to 1, with scores near 1 indicating higher relevance to the question.
On the other hand, ChatGPT identifies incorrect responses and aggregates the correct ones into a cohesive binary answer. While the answer retriever can work without ChatGPT, using them together produces better-quality answers. ChatGPT is more intelligent and better than the adopted answer retriever at detecting incorrect answers. 

We used the ChatGPT version based on the GPT-4 architecture,\footnote{The version of GPT-4 used is 0613, a snapshot of GPT-4 from June 13th 2023.} while the specific prompt template we used is:\smallskip

\begin{quoting}[font=itshape]
\noindent Output a comprehensive answer based only and exclusively on the information within the paragraphs below (if any can be used to answer) which were extracted from the documentation to be assessed. If no paragraph can answer the question, then output only "No, I cannot answer". Otherwise, the comprehensive answer must contain citations to the source paragraphs, e.g., blablabla (paragraphs 1 and 2), blabla (paragraph 0). It should also start with "Yes" if the answer is positive, "No" if the answer is negative, or "N/A" if the answer is not available. \\
Question: \{question\} \\
Paragraphs: \{contents\}
\end{quoting}\smallskip
	
This template is designed for re-elaborating the output of an automated answer retrieval system like the one we employed. The prompt consists of guidelines on how to form a comprehensive answer based on the paragraphs provided for a specific question. It asks the system to generate an answer solely based on the information found in these paragraphs. It also requests that the source paragraphs be cited in the final answer for transparency. If no paragraph can answer the question, then the system is instructed to output \quotes{No, I cannot answer}.

To further dissect the instructions:
\begin{itemize}
    \item \quotes{Output a comprehensive answer based only and exclusively on the information within the paragraphs below} means the algorithm should strictly use the provided text.
    \item \quotes{if any can be used to answer} directs the algorithm to check the relevancy of the provided text snippets.
    \item \quotes{If no paragraph can answer the question, then output only 'No, I cannot answer'} serves as a guideline for cases where the snippets do not have the required data.
    \item The directive that the answer \quotes{must contain citations to the source paragraphs} mandates the algorithm to reference its sources, ensuring transparency and reliability.
\end{itemize}

The phrase \quotes{extracted from the documentation to be assessed} is a key instruction aimed at clarifying the scope of where the answer should come from. This instruction is particularly necessary for cases where the question might have an implicit answer within the given documentation, but not an explicit one. 
In other words, it helps ChatGPT make an inference based on the information that is available, rather than stating \quotes{I cannot answer} simply because an explicit answer is not given.

Once this filtration and aggregation process is complete, we apply the DoX metric to assess the explanatory depth of these refined answers. 


The culmination of this process is the \textit{explanatory relevance score}, which is calculated using the formula: \[ \text{Explanatory Relevance Score} = \text{DoX Score} \times \text{Max(Pertinence Score)} \]
Here, \quotes{Max(Pertinence Score)} is essentially the highest pertinence score among the retrieved correct answers for a checklist question. The explanatory relevance score thus stands as an aggregate metric that captures both the depth of explanations and the specificity of content in terms of compliance with the regulations embodied in the checklist considered.

\section{Results} \label{sec:results}


As suggested by the qualitative analysis presented in \Cref{sec:case_study_n_checklist}, platforms may exhibit varying degrees of compliance regarding ranking transparency. To verify this hypothesis, we went through a manual assessment of the six selected platforms' documentation, seeking to understand how well they comply with the level of detail and completeness of information required by the \ac{P2B} Regulation. Then, we compared the results from the manual assessment with those obtained by using the automated assessment tools presented in \Cref{sec:algorithms}.

For our manual evaluation, we employed two distinct methods.
The first approach involved the expertise of three legal professionals who reviewed all relevant documentation and checklist inquiries. 
Conversely, our second method engaged over 130 participants. Given the associated costs, this method was applied to just two platforms and was narrowed to only four key questions.

The data mentioned in this paper and the code used for this experiment are available at \cite{replication_package}.

\subsection{Comprehensive Review by Legal Experts} \label{sec:results:experts}


Three Belgian senior legal experts each independently rated the documentation of the six platforms identified using the checklist from Section \ref{sec:case_study_n_checklist:checklist}. Among these experts, two are academic researchers specializing in EU law on information and communication technologies, with respectively 7 and 5 years of experience in this field. The third, also well-versed in the same domain, with 6 years of experience, is a co-author of our study. To avoid biased results from this legal expert, as well as from the other two, three mitigation measures were taken. First, the set of questions of the checklist was elaborated objectively, before the evaluation of platforms' documentation. Second, the scores attributed to platforms' documentation during the assessments were based on an objective scale, described below. Third, legal experts compared and discussed their individual assessments after they were collected, to verify whether bias or other significant discrepancies appeared in the results of the co-author of the study, which was not the case. 

These experts were instructed to answer the checklist exclusively based on the link sets identified in \Cref{sec:case_study_n_checklist:data_collection}. \Cref{tab:platform_stats} provides more details on the size of these link sets.

The experts were requested to use a scale from 1 to 5, defined as follows:
\begin{itemize}
    \item \textbf{1}: The question is not answered at all.
    \item \textbf{2}: Indirectly or very poorly answered.
    \item \textbf{3}: Partly answered.
    \item \textbf{4}: Quite good, but not fully sufficient \emph{vis à vis} the legal standard.
    \item \textbf{5}: Satisfactory -- not necessarily perfect but close to the legal standard.
\end{itemize}
This scoring system allows for easier averaging across both experts and platforms, providing an aggregate measure of compliance.

During this first manual assessment, we rapidly found out that substantial differences existed between several platform operators. 
As illustrated in \Cref{tab:avg_compliance_by_type}, Bing led with an average compliance score of 3.5. It was followed in descending order by Tripadvisor, Amazon, Google, Booking, and Yahoo.\footnote{The low score for Yahoo is attributed to its reliance on Bing's documentation, as our assessment focused solely on Yahoo's own documents.} Given that a score of `2' indicates a vague or indirect response, our findings resonate with the observations made by the European Commission about the quality of these explanations (\Cref{sec:related_work}).

Yahoo registered the smallest average variance in scores (0.12), indicating minimal disagreement among reviewers. Conversely, Bing and Google exhibited the most variance. As per \Cref{tab:platform_stats}, Yahoo and Bing/Google respectively had the fewest and most links under consideration, one might deduce that the volume of documentation influences expert ratings. This makes intuitive sense, as processing and retaining information from extensive texts can be challenging.

\begin{table}[htb]
    \centering
    \caption{Experts' Assessment: Average compliance scores given by the three experts for each platform, grouped by type and sorted by score. The scores are averaged across all the experts and all the questions for each platform.}
    \label{tab:avg_compliance_by_type}
    \begin{tabular}{llr} 
        Type & Platform & \multicolumn{1}{l}{Average Compliance Score} \\ 
        \midrule
        \multirow{3}{*}{\textit{Intermediary Service}} & TripAdvisor & 2.627 \\
        & Amazon & 2.569 \\
        & Booking & 1.745 \\ 
        \midrule
        \multirow{3}{*}{\textit{Search Engine}} & Bing & 3.544 \\
        & Google & 2.088 \\
        & Yahoo & 1.228
    \end{tabular}
\end{table}

Relying on only three legal experts, albeit realistic given the expense of specialists in the \ac{P2B} Regulation, could skew interpretations. The small sample size of experts might not cover the breadth of possible scenarios or the diversity of people's opinions.
Therefore, to mitigate this issue and corroborate the preliminary insights with more empirical data, we initiated a broader empirical study, focusing on evaluating software documentation from leading online intermediation services, namely Tripadvisor and Booking. 

\subsection{Large-scale Manual Assessment} \label{sec:results:prolific}


Our larger scale empirical study involved 134 non-expert participants, sourced through Prolific.\footnote{\url{https://www.prolific.co}} Following the same methodology as the study conducted with experts (\Cref{sec:results:experts}), participants were asked to evaluate the documentation identified for Tripadvisor and Booking presented in \Cref{sec:case_study_n_checklist:data_collection}. They rated, on a 1-5 scale, how effectively the documentation answered questions related to the explanatory information mandated by the \ac{P2B} regulations.

Guided by the European Commission guidelines (\Cref{sec:background:p2b_regulation}) stating that explanations about the ranking system should cater to the \quotes{average} users' technical aptitude and needs for a given service (\Cref{sec:background:p2b_regulation}), participants were instructed to adopt the perspective of business users employing online intermediation services or online search engines.

To maintain the study's feasibility and ensure a median completion within approximately 10 minutes, we considered only four of the 17 questions presented in \Cref{sec:case_study_n_checklist} and limited the analysis to the documentation of two intermediation services: Booking and Tripadvisor.  These were chosen based on the brevity of their explanations (Booking is seven webpages long, Tripadvisor nine) and were representative of the best and worst explanations among the selected set of online intermediation services, as adjudged by the legal experts. 

We curated the four questions, drawing from the list in \Cref{sec:case_study_n_checklist}, by considering the objective requirements of the regulation and its main goal, i.e., being able to understand how the ranking works and understanding how to improve/change the results of the ranking to improve outcomes.

Eventually, the selected questions, modified to fit a 1 (strongly disagree) to 5 (strongly agree) scale, were:
\begin{enumerate}[Q1]
    \setcounter{enumi}{2}
    \item \textbf{Clarity of Ranking Mechanism:} The documentation clearly and sufficiently explains the mechanics and main parameters of the ranking system. Do you agree?
    \setcounter{enumi}{3}
    \item \textbf{Rationale for Ranking Parameters:} The documentation satisfactorily clarifies why certain parameters are critical in determining ranking instead of others. Do you agree?
    \setcounter{enumi}{14}
    \item \textbf{Improving Ranking:} The documentation adequately instructs users on how to improve the ranking of their goods or services. Do you agree?
    \setcounter{enumi}{15}
    \item \textbf{Altering Ranking Through Payments:} The documentation transparently discloses the paid options to improve ranking, and the effects thereof. Do you agree?
\end{enumerate}

\begin{figure}
    \centering
    \includegraphics[width=.9\linewidth]{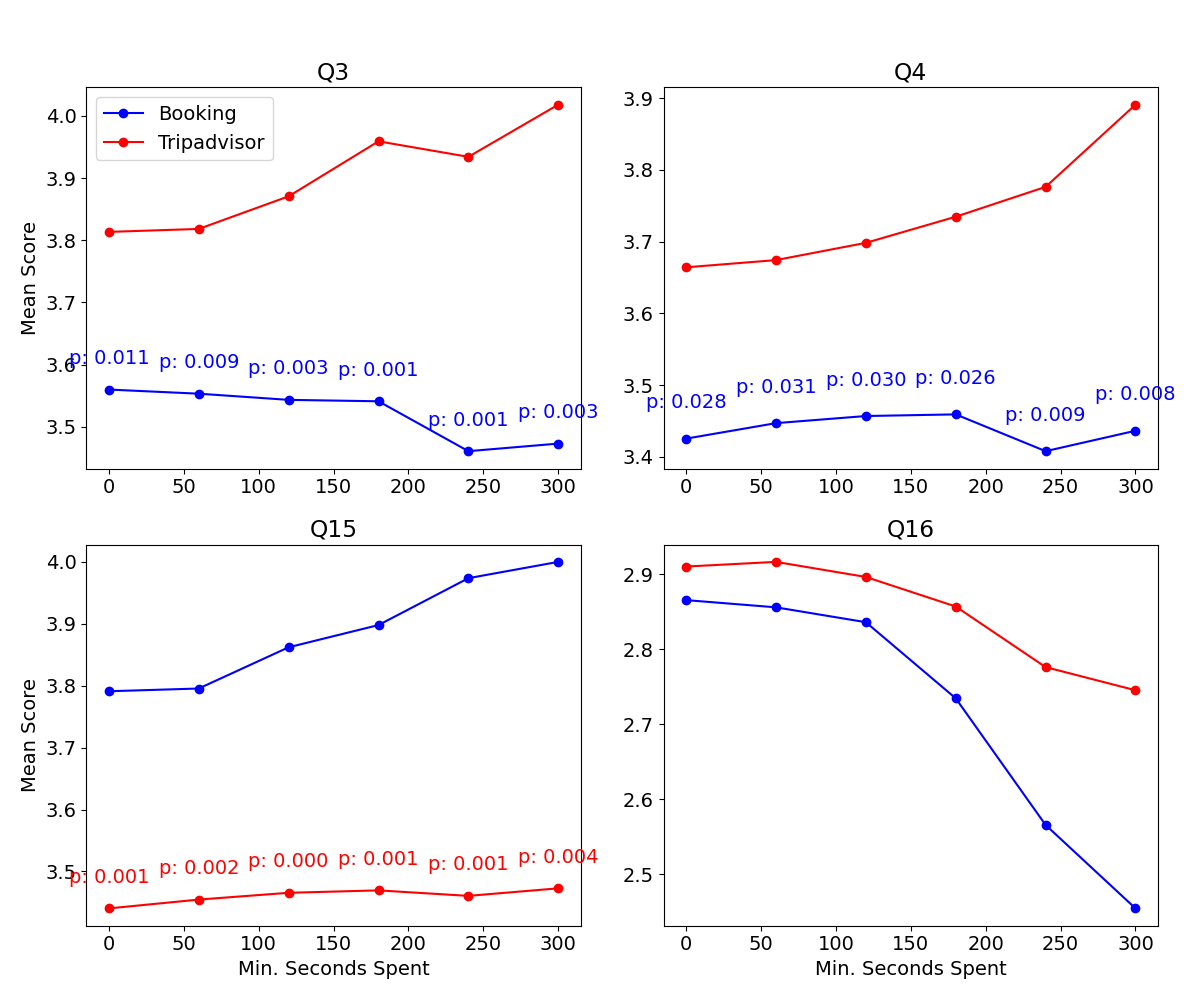}
    \caption{Average agreement rates for participants reviewing documentation on Tripadvisor and Booking, considering review durations from five to zero minutes. Includes statistically significant p-values for agreement rate comparisons between platforms.}
    \label{fig:means_when_changing_thresholds}
\end{figure}

The study was designed to have evenly distributed male and female participants.
Participants were only allowed from the UK\footnote{Since after Brexit, the UK has the discretion to retain, modify, or discard the provisions of the \ac{P2B} Regulation.} or Ireland which are the only English-speaking countries where the \ac{P2B} Regulation applies.
We applied further pre-screening on the spoken language (participants had to be fluent in English) and the Prolific approval rate \cite{eyal2021data} which had to be above 75\%.

Eventually, the study results aligned with our initial expectations, with observed differences between the perceived quality of explanations in the documentation of Tripadvisor and Booking. Furthermore, a few one-sided Mann-Whitney U tests also confirmed that these differences are statistically significant in most cases, as summarized in Figure~\ref{fig:means_when_changing_thresholds}:
\begin{itemize}
    \item Tripadvisor surpasses Booking in detailing the ranking mechanism (Q\ref{q3}) and its rationale (Q\ref{q4}). The effect sizes for these observations are 0.42 and 0.44, respectively, suggesting a moderately strong relationship.
    \item Conversely, Tripadvisor lags behind Booking when it comes to guidance on enhancing rankings (Q\ref{q15}). This observation has a notably larger effect size of 0.60, indicating a stronger relationship.
    \item Regarding the transparency of paid ranking improvement options (Q\ref{q16}), the effect size is 0.49, suggesting a moderate relationship. Although no significant difference was measured, Tripadvisor still slightly outperforms Booking.
\end{itemize}

As shown in \Cref{fig:means_when_changing_thresholds}, a deeper engagement with the documentation (achieved by spending more minutes) highlights the differences between Booking and Tripadvisor. That is because, intuitively, the longer one spends reading the documents, the more likely they are to notice nuances and differences. 

However, only 55 of the 134 participants dedicated more than 5 minutes to reading the documentation.
This points to the importance of providing brief and clear summaries at the start of documentation for quick information access. This would ensure that the most relevant information is efficiently communicated to non-expert users, within a reasonably limited timeframe.

The data in \Cref{fig:means_when_changing_thresholds} closely align with the experts' evaluations for each question. For instance, for Q3, experts ranked Tripadvisor at an average of 4.66, a score significantly higher than Booking's 2.33.
In Q4, experts assigned Tripadvisor a rating of 3.66, outperforming Booking's 1.66.
During Q15's assessment, Booking garnered a score of 4, marginally edging out Tripadvisor's 3.33.
For Q16, experts gave Tripadvisor a 2.66, with Booking trailing at 1.

By cross-referencing the figures and experts' scores, there is a consistent pattern between the two, reinforcing the validity and reliability of the study findings.

Importantly, these findings were not only statistically significant but their effect sizes were medium to large, according to Cohen's conventional criteria. In particular, the effect sizes for Q\ref{q3} and Q\ref{q4} demonstrate a moderate practical advantage for Tripadvisor. While a large effect size for Q\ref{q15} underlines the shortcomings of Tripadvisor when compared to Booking.

\begin{table}[ht]
    \centering
    \caption{Comparative Statistical Data: Booking vs. Tripadvisor. Average scores are shown with their standard deviation. For each question, the best scores are shown in bold.}
    \label{tab:raw_stats}
        \begin{tabular}{|l|c|c|} 
            \hhline{~--|}
            \multicolumn{1}{l|}{}                                                                                                                                    & {\cellcolor[rgb]{0.898,0.898,0.898}}Booking & {\cellcolor[rgb]{0.898,0.898,0.898}}Tripadvisor  \\ 
            \hline
            {\cellcolor[rgb]{0.898,0.898,0.898}}\begin{tabular}[c]{@{}>{\cellcolor[rgb]{0.898,0.898,0.898}}l@{}}Q\ref{q3}: Clarity of~\\Ranking Mechanism\end{tabular}      & 3.56 ± 0.95                                 & \textbf{3.81 ± 0.86}                             \\ 
            \hline
            {\cellcolor[rgb]{0.898,0.898,0.898}}\begin{tabular}[c]{@{}>{\cellcolor[rgb]{0.898,0.898,0.898}}l@{}}Q\ref{q4}: Rationale for \\Ranking Parameters\end{tabular}  & 3.43 ± 1.04                                 & \textbf{3.66 ± 0.96}                             \\ 
            \hline
            {\cellcolor[rgb]{0.898,0.898,0.898}}Q\ref{q15}: Improving Ranking                                                                                                & \textbf{3.79 ± 0.89}                        & 3.44 ± 0.91                                      \\ 
            \hline
            {\cellcolor[rgb]{0.898,0.898,0.898}}\begin{tabular}[c]{@{}>{\cellcolor[rgb]{0.898,0.898,0.898}}l@{}}Q\ref{q16}: Altering Ranking~\\Through Payments\end{tabular} & 2.87 ± 1.17                                 & \textbf{2.91 ± 1.24}                             \\
            \hline
        \end{tabular}
\end{table}

\subsection{Automated Assessments} \label{sec:results:automated}

As shown so far, online platforms often struggle to produce clear documentation that complies with regulations like \ac{P2B}. Evaluating and ensuring the quality of such documentation is challenging, largely due to the lack of standardized automated tools for assessing compliance and explainability across different platform providers.

To address this, we employed the assessment tools detailed in Section \ref{sec:algorithms}. Our aim was to determine if these tools' evaluations are consistent with expert opinions and the empirical findings discussed in Sections \ref{sec:results:experts} and \ref{sec:results:prolific}. We also compared their performance against baseline methods such as constant "Yes" responses and random\footnote{We used 42 as a random seed.} assessments.

The findings in Table \ref{tab:experts_vs_tools} show that the DoX-based method and ChatGPT 3.5 match experts' evaluations 63.88\% of the time. However, ChatGPT 4 is the least effective AI tool, performing even worse than random assessment. Since ChatGPT is a black-box, understanding why this occurs is impossible to us.


The scoring between the experts and ChatGPT was normalized as follows: any response scoring 3 or higher (indicating the question is ``partially answered", as detailed in \Cref{sec:results:experts}) was categorized as `Yes', while scores below 3 were labeled `No'. The DoX-based method already provides a straightforward `Yes' or `No' response.

\begin{table}[htb]
    \centering
    \caption{Agreements Rates: Percentage of outputs generated by different automated assessment strategies that align with that of the experts (\Cref{sec:results:experts}). For example, for Bing, ChatGPT 3.5 agreed with the majority of experts' answers 68.42\% of the time. Scores in bold are the highest row-wise.} \label{tab:experts_vs_tools}
    \resizebox{\linewidth}{!}{
        \begin{tabular}{l|r|r|r|r|r}
            Platform & ChatGPT 3.5 & ChatGPT 4 & DoX-based & Random & Always Yes \\ \hline
            Bing       & 68.42\%      & 31.58\%     & 68.42\%     & 57.89\%  & \textbf{73.68\%}      \\
            Booking    & \textbf{76.47\%}      & 29.41\%     & 35.29\%     & 35.29\%  & 5.88\%       \\
            Tripadvisor& 58.82\%      & 52.94\%     & \textbf{64.71\%}     & 41.18\%  & 58.82\%      \\
            Amazon     & 47.06\%      & 52.94\%     & \textbf{64.71\%}     & 47.06\%  & \textbf{64.71\%}      \\
            Google     & 36.84\%      & \textbf{57.89\%}     & 52.63\%     & \textbf{57.89\%}  & 31.58\%      \\
            Yahoo      & \textbf{94.74\%}      & \textbf{94.74\%}     & \textbf{94.74\%}     & 42.11\%  & 5.26\%       \\ \hline
            All & \textbf{63.89\%}      & 53.70\%     & \textbf{63.89\%}     & 54.63\%  & 39.81\%      \\
        \end{tabular}
    }
\end{table}

We also performed Mann-Whitney U tests and a rank biserial correlation analysis to analyze the alignment between the explanatory relevance, pertinence, and DoX scores produced by the second tool (\Cref{sec:algorithms:dox}) and the yes-or-no experts' answers. 

\begin{figure}[htb]
    \centering
    \includegraphics[width=.9\linewidth]{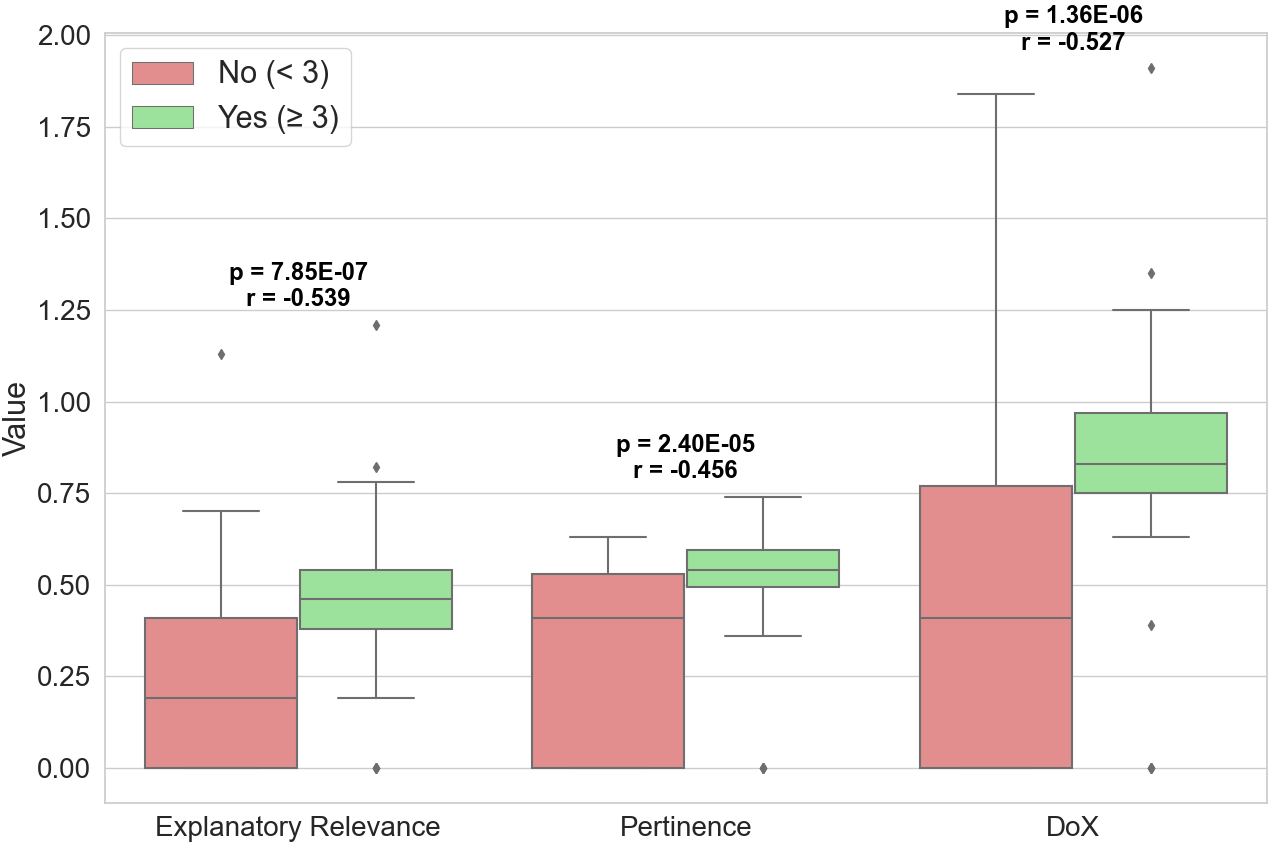}
    \caption{Boxplot comparing `Explanatory Relevance', `Pertinence', and `DoX' by experts' answers. Annotations indicate \( p \)-values and rank biserial correlation (\( r \)).}
    \label{fig:scores_alignment}
\end{figure}

As shown in \Cref{fig:scores_alignment}, all three scores show a significant \( p \)-value (less than .05), implying that the difference between the `Yes' and `No' groups in experts' answers is statistically significant for each score. The rank biserial correlation values range from -.456 to -.539, thus suggesting a moderate to strong negative correlation between the scores and the experts' answers. The explanatory relevance score correlates the best with the majority of answers, indicated by the highest absolute value of the rank biserial correlation (-.539).

\section{Discussion \& Threats to Validity} \label{sec:discussion}

The findings from Sections \ref{sec:results:experts} and \ref{sec:results:prolific} highlight a statistically significant disparity in compliance scores across diverse platforms. This implies varying levels of compliance among platforms regarding ranking transparency. A possible reason for this inconsistency is the absence of standardized automated tools for gauging compliance and explainability.

Section \ref{sec:results:automated} implies that automated tools could offer a solution. Notably, results indicate that ChatGPT 3.5 and the DoX-based approach, the best ones, align with the majority of experts 63.88\% of the time. However, they are far from flawless, indicating there is considerable potential for improvement. 

Delving deeper, the DoX-based approach seems to perform better than GPT-3.5 when evaluating longer documents. As illustrated in Table \ref{tab:platform_stats}, for documentation from Google (52 docs, averaging 1679 words each), Bing (16 docs, averaging 964 words each), and Tripadvisor (10 pages, averaging 1653 words each), the DoX-based method achieves agreement rates of 52.63\%, 68.42\%, and 64.71\%, respectively. In contrast, GPT-3.5 scores 36.84\%, 68.42\%, and 58.82\%. These results can be attributed to the design of the DoX-based approach, which is optimized to handle long documents effectively, bypassing the input size constraints of ChatGPT.

We used ChatGPT without any fine-tuning. While the DoX-based approach incorporates GPT-4, its performance exceeds that of GPT-4 without answer retrieval. This hints at the instrumental role that prompt engineering might have in bolstering these tools' efficacy.

It is also crucial to understand the nuances of the DoX-based approach. Though this approach aligns with experts 63.88\% of the time, this alignment merely considers ChatGPT's binary (`yes' or `no') responses and omits the explanatory relevance scores. An in-depth examination of the alignment between the explanatory relevance scores from our DoX-based methodology and expert responses reveals their potential in pinpointing documentation sections that are non-compliant or inadequately articulated. \Cref{fig:scores_alignment} illustrates that lower scores correlate with expert disapproval. Thus, the feedback depth from the DoX-based approach exceeds that of ChatGPT.


We acknowledge a potential limitation of this study: as discussed in \Cref{sec:case_study_n_checklist:data_collection}, the study focused solely on textual content. However, some platforms also feature video content or imagery that could provide additional insights. Yet, for intermediation services at least, it is compulsory to provide their explanations in written form (\Cref{sec:case_study_n_checklist:data_collection}), hence our decision to focus on textual content. We cannot rule out that information not available in textual content could be embedded in videos or images.


Given the performance scores, the fully automated tools should not replace legal experts. These tools are effective at tracking flaws in documentation, but their best use is with legal professionals. Automated tools could be used to determine if existing documentation shows insufficient compliance, and if so, whether a legal expert is needed to enhance the problematic documentation. 


Finally, \Cref{tab:experts_vs_tools} highlights cases in which ChatGPT 3.5 is the top performer, and cases in which the DoX-based method is more effective. Thus, for the best results, it seems reasonable to suggest combining both tools, then proceeding with manual validation.



\section{Societal Implications} \label{sec:societal_implications}

The qualitative assessment approaches to compliance measurement, and automated tools that we proposed in the previous sections have various societal implications, which we underline hereafter.


Our study provides concrete data and assessments of major market actors' documentation, in terms of compliance with a specific piece of EU law. It goes beyond existing literature by combining insights from legal experts, laypersons, and automated tools. This research can highlight ranking practices impacting individuals, businesses, and society. It can also assist platforms in understanding their level of compliance compared to competitors and addressing any issues. These results are beneficial for society.

Additionally, the paper highlights that assessing the quality of software documentation is a difficult, resource-consuming task, due to the lack of standardized automated tools to evaluate compliance. The paper also proposes a solution to this gap, in the form of automated assessment tools that are tested against human judgment. The societal benefits of this solution include enhanced means for regulators and public authorities to verify and force compliance with ranking transparency requirements. This, in turn, may lead to more balanced and fair markets for business and corporate website users, as well as more competition between platforms, as intended by the EU legislator. 
These benefits contribute to the United Nations Sustainable Development Goal 10.3 \cite{UN2023}, aimed at reducing outcome inequalities and promoting equal opportunities.

Finally, our research explores approaches that can measure the quality of explanations provided about software and algorithmic processes \textit{vis-à-vis} a legal standard, which is actually a crucial field of research. This is particularly due to the rapid development of \textit{i)} machine learning techniques and tools, \textit{ii)} methods to explain machine learning systems and their outputs \cite{speith2022review}, and \textit{iii)} legal requirements imposing explanations when such systems are used \cite{hacker2020varieties}. In the context of our paper, it is clear that the ranking systems at stake include machine learning technology, which can be difficult to explain to laypersons. As lawmakers impose on online platforms to disclose information about these systems, it is particularly beneficial to society to be able to measure the quality of the explanations provided and determine when an explanation is insufficient or of poor quality. Besides, this reasoning and the tools that we develop can be transposed to other contexts implying the compulsory provision of explanations in relation to machine learning tools or outputs, such as the forthcoming AI Act (see, for instance, Art. 13 of the proposed AI Act \cite{EC2021AIAct}). 

\section{Conclusions \& Future Directions} \label{sec:conclusion}

The EU's \ac{P2B} Regulation requires online platforms to clearly explain their ranking procedures. In this context, we studied how platforms like Amazon, Booking, Google, Bing, Tripadvisor, and Yahoo comply with this regulation.

With three legal experts and feedback from 130 users, we analyzed the documentation. Our findings showed varying compliance levels. One reason might be the lack of standardized assessment criteria or automated tools.

To address this, we introduced two automated assessment tools. First, given its popularity, we considered ChatGPT as an assessment method. Next, we proposed a hybrid method (designed to better handle long documentation) merging ChatGPT with answer retrieval techniques and a new metric, DoX, to measure documentation explainability.

When we compared the results from our automated tools to manual assessments, they agreed with human evaluations 63.88\% of the time. This suggests these tools could help make the review process more efficient. Importantly, the method based on DoX was particularly good at assessing lengthy documents. Furthermore, its explanatory scores aligned well with expert opinions, implying that content with weak explanations tends to receive lower compliance ratings from professionals.

While these tools have limitations, their continuous assessment capability can be invaluable for regulators and the public. Combining them with expert judgment can promote transparent documentation, essential for a fair digital business space. This transparency aligns with goals like the UN's Sustainable Development Goal 10.3, fostering an informed and equitable online ecosystem.

Future improvements could refine these tools through better prompt engineering, tailoring models like ChatGPT for compliance tasks, and including multimedia analysis capabilities.

\begin{acks}
F. Sovrano and A. Bacchelli acknowledge the support of the Swiss National Science Foundation for the SNF Project 200021\_197227. M. Lognoul aknowledges the support of the European Union H2020 research and innovation program, Grant Agreement No. 958339 (DENiM).
\end{acks}

\section*{Author Contributions}
F. Sovrano and M. Lognoul: conceptualization, methodology, data curation, original draft preparation, investigation, validation, formal analysis. 
F. Sovrano: software, visualization.
A. Bacchelli: methodology conceptualization, review, and editing.

\newpage
\bibliographystyle{ACM-Reference-Format}
\bibliography{references}


\begin{thebibliography}{35}


\ifx \showCODEN    \undefined \def \showCODEN     #1{\unskip}     \fi
\ifx \showDOI      \undefined \def \showDOI       #1{#1}\fi
\ifx \showISBNx    \undefined \def \showISBNx     #1{\unskip}     \fi
\ifx \showISBNxiii \undefined \def \showISBNxiii  #1{\unskip}     \fi
\ifx \showISSN     \undefined \def \showISSN      #1{\unskip}     \fi
\ifx \showLCCN     \undefined \def \showLCCN      #1{\unskip}     \fi
\ifx \shownote     \undefined \def \shownote      #1{#1}          \fi
\ifx \showarticletitle \undefined \def \showarticletitle #1{#1}   \fi
\ifx \showURL      \undefined \def \showURL       {\relax}        \fi
\providecommand\bibfield[2]{#2}
\providecommand\bibinfo[2]{#2}
\providecommand\natexlab[1]{#1}
\providecommand\showeprint[2][]{arXiv:#2}

\bibitem[Achinstein(2010)]%
        {achinstein2010evidence}
\bibfield{author}{\bibinfo{person}{Peter Achinstein}.}
  \bibinfo{year}{2010}\natexlab{}.
\newblock \bibinfo{booktitle}{\emph{Evidence, explanation, and realism: Essays
  in philosophy of science}}.
\newblock \bibinfo{publisher}{Oxford University Press}.
\newblock
\showISBNx{978-0199735259}


\bibitem[Alkaissi and McFarlane(2023)]%
        {alkaissi2023artificial}
\bibfield{author}{\bibinfo{person}{Hussam Alkaissi} {and}
  \bibinfo{person}{Samy~I McFarlane}.} \bibinfo{year}{2023}\natexlab{}.
\newblock \showarticletitle{Artificial hallucinations in ChatGPT: implications
  in scientific writing}.
\newblock \bibinfo{journal}{\emph{Cureus}} \bibinfo{volume}{15},
  \bibinfo{number}{2} (\bibinfo{year}{2023}).
\newblock


\bibitem[Azamfirei et~al\mbox{.}(2023)]%
        {azamfirei_large_2023}
\bibfield{author}{\bibinfo{person}{Razvan Azamfirei}, \bibinfo{person}{Sapna~R.
  Kudchadkar}, {and} \bibinfo{person}{James Fackler}.}
  \bibinfo{year}{2023}\natexlab{}.
\newblock \showarticletitle{Large language models and the perils of their
  hallucinations}.
\newblock \bibinfo{journal}{\emph{Critical Care}} \bibinfo{volume}{27},
  \bibinfo{number}{1} (\bibinfo{date}{March} \bibinfo{year}{2023}),
  \bibinfo{pages}{120}.
\newblock
\showISSN{1364-8535}


\bibitem[Bibal et~al\mbox{.}(2021)]%
        {bibal2021legal}
\bibfield{author}{\bibinfo{person}{Adrien Bibal}, \bibinfo{person}{Michael
  Lognoul}, \bibinfo{person}{Alexandre De~Streel}, {and}
  \bibinfo{person}{Beno{\^\i}t Fr{\'e}nay}.} \bibinfo{year}{2021}\natexlab{}.
\newblock \showarticletitle{Legal requirements on explainability in machine
  learning}.
\newblock \bibinfo{journal}{\emph{Artificial Intelligence and Law}}
  \bibinfo{volume}{29} (\bibinfo{year}{2021}), \bibinfo{pages}{149--169}.
\newblock


\bibitem[Bowman et~al\mbox{.}(2015)]%
        {bowman2015large}
\bibfield{author}{\bibinfo{person}{Samuel~R. Bowman}, \bibinfo{person}{Gabor
  Angeli}, \bibinfo{person}{Christopher Potts}, {and}
  \bibinfo{person}{Christopher~D. Manning}.} \bibinfo{year}{2015}\natexlab{}.
\newblock \showarticletitle{A large annotated corpus for learning natural
  language inference}. In \bibinfo{booktitle}{\emph{Proceedings of the 2015
  Conference on Empirical Methods in Natural Language Processing, {EMNLP} 2015,
  Lisbon, Portugal, September 17-21, 2015}},
  \bibfield{editor}{\bibinfo{person}{Llu{\'{\i}}s M{\`{a}}rquez},
  \bibinfo{person}{Chris Callison{-}Burch}, \bibinfo{person}{Jian Su},
  \bibinfo{person}{Daniele Pighin}, {and} \bibinfo{person}{Yuval Marton}}
  (Eds.). \bibinfo{publisher}{The Association for Computational Linguistics},
  \bibinfo{pages}{632--642}.
\newblock
\urldef\tempurl%
\url{https://doi.org/10.18653/v1/d15-1075}
\showURL{%
\tempurl}


\bibitem[Brynjolfsson and Kahin(2002)]%
        {brynjolfsson2002understanding}
\bibfield{author}{\bibinfo{person}{Erik Brynjolfsson} {and}
  \bibinfo{person}{Brian Kahin}.} \bibinfo{year}{2002}\natexlab{}.
\newblock \bibinfo{booktitle}{\emph{Understanding the digital economy: data,
  tools, and research}}.
\newblock \bibinfo{publisher}{MIT press}.
\newblock


\bibitem[Brynjolfsson and Oh(2012)]%
        {DBLP:conf/icis/BrynjolfssonO12}
\bibfield{author}{\bibinfo{person}{Erik Brynjolfsson} {and}
  \bibinfo{person}{JooHee Oh}.} \bibinfo{year}{2012}\natexlab{}.
\newblock \showarticletitle{The Attention Economy: Measuring the Value of Free
  Digital Services on the Internet}. In \bibinfo{booktitle}{\emph{Proceedings
  of the International Conference on Information Systems, {ICIS} 2012, Orlando,
  Florida, USA, December 16-19, 2012}}. \bibinfo{publisher}{Association for
  Information Systems}.
\newblock
\urldef\tempurl%
\url{http://aisel.aisnet.org/icis2012/proceedings/EconomicsValue/9}
\showURL{%
\tempurl}


\bibitem[Busch(2023)]%
        {busch2023algorithmic}
\bibfield{author}{\bibinfo{person}{Christoph Busch}.}
  \bibinfo{year}{2023}\natexlab{}.
\newblock \showarticletitle{From Algorithmic Transparency to Algorithmic
  Choice: European Perspectives on Recommender Systems and Platform
  Regulation}.
\newblock In \bibinfo{booktitle}{\emph{Recommender Systems: Legal and Ethical
  Issues}}. \bibinfo{publisher}{Springer International Publishing Cham},
  \bibinfo{pages}{31--54}.
\newblock


\bibitem[Commission(2023)]%
        {EU2023P2B}
\bibfield{author}{\bibinfo{person}{European Commission}.}
  \bibinfo{year}{2023}\natexlab{}.
\newblock \bibinfo{title}{Report from the Commission on the first preliminary
  review of the P2B Regulation}.
\newblock
\newblock
\urldef\tempurl%
\url{https://digital-strategy.ec.europa.eu/en/library/report-commission-first-preliminary-review-p2b-regulation}
\showURL{%
\tempurl}
\newblock
\shownote{Accessed on: 2023-10-06}.


\bibitem[European~Commission and Technology(2020)]%
        {EC2020P2BGuidelines}
\bibfield{author}{\bibinfo{person}{Content European~Commission,
  Directorate-General for Communications~Networks} {and}
  \bibinfo{person}{Technology}.} \bibinfo{year}{2020}\natexlab{}.
\newblock \bibinfo{title}{Commission Notice Guidelines on ranking transparency
  pursuant to Regulation (EU) 2019/1150 of the European Parliament and of the
  Council 2020/C 424/01}.
\newblock \bibinfo{howpublished}{EUR-Lex}.
\newblock
\urldef\tempurl%
\url{https://eur-lex.europa.eu/legal-content/EN/TXT/?uri=CELEX:52020XC1208(01)}
\showURL{%
\tempurl}
\newblock
\shownote{Accessed on: 2023-10-06}.


\bibitem[European~Commission and Technology(2021)]%
        {EC2021AIAct}
\bibfield{author}{\bibinfo{person}{Content European~Commission,
  Directorate-General for Communications~Networks} {and}
  \bibinfo{person}{Technology}.} \bibinfo{year}{2021}\natexlab{}.
\newblock \bibinfo{title}{Proposal for a Regulation of the European Parliament
  and of the Council laying down harmonised rules on Artificial Intelligence
  (Artificial Intelligence Act) and amending certain Union legislative acts}.
\newblock
  \bibinfo{howpublished}{\url{https://eur-lex.europa.eu/legal-content/EN/TXT/?uri=celex:52021PC0206}}.
\newblock
\newblock
\shownote{Accessed on: 2023-10-06}.


\bibitem[{European Parliament and Council}(2019)]%
        {eurlex2019}
\bibfield{author}{\bibinfo{person}{{European Parliament and Council}}.}
  \bibinfo{year}{2019}\natexlab{}.
\newblock \bibinfo{title}{Regulation (EU) 2019/1150 of the European Parliament
  and of the Council of 20 June 2019 on promoting fairness and transparency for
  business users of online intermediation services (Text with EEA relevance)}.
\newblock
\newblock
\urldef\tempurl%
\url{https://eur-lex.europa.eu/legal-content/EN/TXT/?uri=CELEX:32019R1150}
\showURL{%
\tempurl}
\newblock
\shownote{PE/56/2019/REV/1, OJ L 186, 11.7.2019, p. 57–79}.


\bibitem[Eyal et~al\mbox{.}(2021)]%
        {eyal2021data}
\bibfield{author}{\bibinfo{person}{Peer Eyal}, \bibinfo{person}{Rothschild
  David}, \bibinfo{person}{Gordon Andrew}, \bibinfo{person}{Evernden Zak},
  {and} \bibinfo{person}{Damer Ekaterina}.} \bibinfo{year}{2021}\natexlab{}.
\newblock \showarticletitle{Data quality of platforms and panels for online
  behavioral research}.
\newblock \bibinfo{journal}{\emph{Behavior Research Methods}}
  (\bibinfo{year}{2021}), \bibinfo{pages}{1--20}.
\newblock


\bibitem[Fereday and Muir-Cochrane(2006)]%
        {fereday2006demonstrating}
\bibfield{author}{\bibinfo{person}{Jennifer Fereday} {and}
  \bibinfo{person}{Eimear Muir-Cochrane}.} \bibinfo{year}{2006}\natexlab{}.
\newblock \showarticletitle{Demonstrating rigor using thematic analysis: A
  hybrid approach of inductive and deductive coding and theme development}.
\newblock \bibinfo{journal}{\emph{International journal of qualitative
  methods}} \bibinfo{volume}{5}, \bibinfo{number}{1} (\bibinfo{year}{2006}),
  \bibinfo{pages}{80--92}.
\newblock


\bibitem[Gajardo and Paz(2019)]%
        {DBLP:conf/amcis/GajardoP19}
\bibfield{author}{\bibinfo{person}{Pablo~A. Gajardo} {and}
  \bibinfo{person}{Ariel I.~La Paz}.} \bibinfo{year}{2019}\natexlab{}.
\newblock \showarticletitle{Measuring the strategic business and {IT} alignment
  in a digitally revolutionized economy}. In \bibinfo{booktitle}{\emph{25th
  Americas Conference on Information Systems, {AMCIS} 2019, Canc{\'{u}}n,
  Mexico, August 15-17, 2019}}. \bibinfo{publisher}{Association for Information
  Systems}.
\newblock
\urldef\tempurl%
\url{https://aisel.aisnet.org/amcis2019/org\_transformation\_is/org\_transformation\_is/19}
\showURL{%
\tempurl}


\bibitem[Garousi et~al\mbox{.}(2013)]%
        {DBLP:conf/ease/GarousiGMRS13}
\bibfield{author}{\bibinfo{person}{Golara Garousi}, \bibinfo{person}{Vahid
  Garousi}, \bibinfo{person}{Mahmood Moussavi}, \bibinfo{person}{G{\"{u}}nther
  Ruhe}, {and} \bibinfo{person}{Brian Smith}.} \bibinfo{year}{2013}\natexlab{}.
\newblock \showarticletitle{Evaluating usage and quality of technical software
  documentation: an empirical study}. In \bibinfo{booktitle}{\emph{17th
  International Conference on Evaluation and Assessment in Software
  Engineering, {EASE} '13, Porto de Galinhas, Brazil, April 14-16, 2013}},
  \bibfield{editor}{\bibinfo{person}{Fabio Q.~B. da~Silva},
  \bibinfo{person}{Natalia~Juristo Juzgado}, {and}
  \bibinfo{person}{Guilherme~Horta Travassos}} (Eds.).
  \bibinfo{publisher}{{ACM}}, \bibinfo{pages}{24--35}.
\newblock
\urldef\tempurl%
\url{https://doi.org/10.1145/2460999.2461003}
\showDOI{\tempurl}


\bibitem[Garousi et~al\mbox{.}(2015)]%
        {DBLP:journals/infsof/GarousiGRZMS15}
\bibfield{author}{\bibinfo{person}{Golara Garousi}, \bibinfo{person}{Vahid
  Garousi{-}Yusifoglu}, \bibinfo{person}{G{\"{u}}nther Ruhe},
  \bibinfo{person}{Junji Zhi}, \bibinfo{person}{Mahmood Moussavi}, {and}
  \bibinfo{person}{Brian Smith}.} \bibinfo{year}{2015}\natexlab{}.
\newblock \showarticletitle{Usage and usefulness of technical software
  documentation: An industrial case study}.
\newblock \bibinfo{journal}{\emph{Inf. Softw. Technol.}}  \bibinfo{volume}{57}
  (\bibinfo{year}{2015}), \bibinfo{pages}{664--682}.
\newblock
\urldef\tempurl%
\url{https://doi.org/10.1016/j.infsof.2014.08.003}
\showDOI{\tempurl}


\bibitem[Hacker et~al\mbox{.}(2022)]%
        {hacker2022regulating}
\bibfield{author}{\bibinfo{person}{Philipp Hacker}, \bibinfo{person}{Johann
  Cordes}, {and} \bibinfo{person}{Janina Rochon}.}
  \bibinfo{year}{2022}\natexlab{}.
\newblock \showarticletitle{Regulating Gatekeeper AI and Data: Transparency,
  Access, and Fairness under the DMA, the GDPR, and beyond}.
\newblock \bibinfo{journal}{\emph{arXiv preprint arXiv:2212.04997}}
  (\bibinfo{year}{2022}).
\newblock


\bibitem[Hacker and Passoth(2020)]%
        {hacker2020varieties}
\bibfield{author}{\bibinfo{person}{Philipp Hacker} {and}
  \bibinfo{person}{Jan-Hendrik Passoth}.} \bibinfo{year}{2020}\natexlab{}.
\newblock \showarticletitle{Varieties of AI Explanations Under the Law. From
  the GDPR to the AIA, and Beyond}. In \bibinfo{booktitle}{\emph{International
  Workshop on Extending Explainable AI Beyond Deep Models and Classifiers}}.
  Springer, \bibinfo{pages}{343--373}.
\newblock


\bibitem[Henrickson and Meroño-Peñuela(2023)]%
        {henrickson_prompting_2023}
\bibfield{author}{\bibinfo{person}{Leah Henrickson} {and}
  \bibinfo{person}{Albert Meroño-Peñuela}.} \bibinfo{year}{2023}\natexlab{}.
\newblock \showarticletitle{Prompting meaning: a hermeneutic approach to
  optimising prompt engineering with {ChatGPT}}.
\newblock \bibinfo{journal}{\emph{AI \& SOCIETY}} (\bibinfo{date}{Sept.}
  \bibinfo{year}{2023}).
\newblock
\showISSN{1435-5655}
\urldef\tempurl%
\url{https://doi.org/10.1007/s00146-023-01752-8}
\showDOI{\tempurl}


\bibitem[Karpukhin et~al\mbox{.}(2020)]%
        {karpukhin2020dense}
\bibfield{author}{\bibinfo{person}{Vladimir Karpukhin}, \bibinfo{person}{Barlas
  Oguz}, \bibinfo{person}{Sewon Min}, \bibinfo{person}{Patrick S.~H. Lewis},
  \bibinfo{person}{Ledell Wu}, \bibinfo{person}{Sergey Edunov},
  \bibinfo{person}{Danqi Chen}, {and} \bibinfo{person}{Wen{-}tau Yih}.}
  \bibinfo{year}{2020}\natexlab{}.
\newblock \showarticletitle{Dense Passage Retrieval for Open-Domain Question
  Answering}. In \bibinfo{booktitle}{\emph{Proceedings of the 2020 Conference
  on Empirical Methods in Natural Language Processing, {EMNLP} 2020, Online,
  November 16-20, 2020}}, \bibfield{editor}{\bibinfo{person}{Bonnie Webber},
  \bibinfo{person}{Trevor Cohn}, \bibinfo{person}{Yulan He}, {and}
  \bibinfo{person}{Yang Liu}} (Eds.). \bibinfo{publisher}{Association for
  Computational Linguistics}, \bibinfo{pages}{6769--6781}.
\newblock
\urldef\tempurl%
\url{https://doi.org/10.18653/v1/2020.emnlp-main.550}
\showURL{%
\tempurl}


\bibitem[Li et~al\mbox{.}(2023)]%
        {DBLP:journals/corr/abs-2305-12962}
\bibfield{author}{\bibinfo{person}{Jiazheng Li}, \bibinfo{person}{Lin Gui},
  \bibinfo{person}{Yuxiang Zhou}, \bibinfo{person}{David West},
  \bibinfo{person}{Cesare Aloisi}, {and} \bibinfo{person}{Yulan He}.}
  \bibinfo{year}{2023}\natexlab{}.
\newblock \showarticletitle{Distilling ChatGPT for Explainable Automated
  Student Answer Assessment}.
\newblock \bibinfo{journal}{\emph{CoRR}}  \bibinfo{volume}{abs/2305.12962}
  (\bibinfo{year}{2023}).
\newblock
\urldef\tempurl%
\url{https://doi.org/10.48550/arXiv.2305.12962}
\showDOI{\tempurl}
\showeprint[arXiv]{2305.12962}


\bibitem[Lukovic(2021)]%
        {lukovic2021information}
\bibfield{author}{\bibinfo{person}{Vesna Lukovic}.}
  \bibinfo{year}{2021}\natexlab{}.
\newblock \showarticletitle{INFORMATION ASYMMETRIES IN ALGORITHMS AT DIGITAL
  PLATFORMS: MOTIVATIONS TO PARTICIPATE AND EU REGULATORY APPROACH}.
\newblock \bibinfo{journal}{\emph{EMAN 2021--Economics \& Management: How to
  Cope with Disrupted Times}} (\bibinfo{year}{2021}), \bibinfo{pages}{167}.
\newblock


\bibitem[Novaes and Reck(2017)]%
        {novaes2017carnapian}
\bibfield{author}{\bibinfo{person}{Catarina~Dutilh Novaes} {and}
  \bibinfo{person}{Erich~H. Reck}.} \bibinfo{year}{2017}\natexlab{}.
\newblock \showarticletitle{Carnapian explication, formalisms as cognitive
  tools, and the paradox of adequate formalization}.
\newblock \bibinfo{journal}{\emph{Synth.}} \bibinfo{volume}{194},
  \bibinfo{number}{1} (\bibinfo{year}{2017}), \bibinfo{pages}{195--215}.
\newblock
\urldef\tempurl%
\url{https://doi.org/10.1007/s11229-015-0816-z}
\showDOI{\tempurl}


\bibitem[of~Economic and Affairs(2023)]%
        {UN2023}
\bibfield{author}{\bibinfo{person}{Department of Economic} {and}
  \bibinfo{person}{Social Affairs}.} \bibinfo{year}{2023}\natexlab{}.
\newblock \bibinfo{title}{Goal 10: Reduce inequality within and among
  countries}.
\newblock
\newblock
\urldef\tempurl%
\url{https://sdgs.un.org/goals/goal10}
\showURL{%
\tempurl}
\newblock
\shownote{Accessed on: 2023-10-06}.


\bibitem[on~the Online Platform~Economy(2021)]%
        {PlatformObservatory2021}
\bibfield{author}{\bibinfo{person}{Observatory on~the Online
  Platform~Economy}.} \bibinfo{year}{2021}\natexlab{}.
\newblock \bibinfo{title}{Study on "Support to the Observatory for the Online
  Platform Economy"}.
\newblock
  \bibinfo{howpublished}{\url{https://platformobservatory.eu/app/uploads/2021/01/P2B-Regulation-monitoring-analysis-January-2021.pdf}}.
\newblock
\newblock
\shownote{Accessed on: 2023-10-06}.


\bibitem[Prastitou-Merdi(2021)]%
        {prastitou2021notion}
\bibfield{author}{\bibinfo{person}{Thalia Prastitou-Merdi}.}
  \bibinfo{year}{2021}\natexlab{}.
\newblock \showarticletitle{The Notion of “Online Intermediation Services”
  Found in the New EU Platform Regulation: Who Is Caught After All?}
\newblock In \bibinfo{booktitle}{\emph{EU Internet Law in the Digital Single
  Market}}. \bibinfo{publisher}{Springer}, \bibinfo{pages}{543--559}.
\newblock


\bibitem[Rao et~al\mbox{.}(2023)]%
        {DBLP:journals/corr/abs-2303-01248}
\bibfield{author}{\bibinfo{person}{Haocong Rao}, \bibinfo{person}{Cyril Leung},
  {and} \bibinfo{person}{Chunyan Miao}.} \bibinfo{year}{2023}\natexlab{}.
\newblock \showarticletitle{Can ChatGPT Assess Human Personalities? {A} General
  Evaluation Framework}.
\newblock \bibinfo{journal}{\emph{CoRR}}  \bibinfo{volume}{abs/2303.01248}
  (\bibinfo{year}{2023}).
\newblock
\urldef\tempurl%
\url{https://doi.org/10.48550/arXiv.2303.01248}
\showDOI{\tempurl}
\showeprint[arXiv]{2303.01248}


\bibitem[{Shaping Europe’s digital future}(2023)]%
        {EUObservatory2023}
\bibfield{author}{\bibinfo{person}{{Shaping Europe’s digital future}}.}
  \bibinfo{year}{2023}\natexlab{}.
\newblock \bibinfo{title}{{EU Observatory on the Online Platform Economy}}.
\newblock
  \bibinfo{howpublished}{\url{https://digital-strategy.ec.europa.eu/en/policies/eu-observatory-online-platform-economy}}.
\newblock
\newblock
\shownote{Accessed on: 2023-10-07}.


\bibitem[Sovrano(2023)]%
        {replication_package}
\bibfield{author}{\bibinfo{person}{Francesco Sovrano}.}
  \bibinfo{year}{2023}\natexlab{}.
\newblock \bibinfo{title}{Replication Package for "Automating Regulatory
  Compliance: An Empirical Study on Ranking Transparency in the Software
  Documentation of EU Online Platforms"}.
\newblock
  \bibinfo{howpublished}{\url{https://github.com/Francesco-Sovrano/Automating-Regulatory-Compliance-An-Empirical-Study-on-Ranking-Transparency-of-EU-Online-Platforms}}.
\newblock
\urldef\tempurl%
\url{https://doi.org/10.5281/zenodo.10478546}
\showDOI{\tempurl}


\bibitem[Sovrano et~al\mbox{.}(2020)]%
        {DBLP:conf/jurix/SovranoPV20}
\bibfield{author}{\bibinfo{person}{Francesco Sovrano}, \bibinfo{person}{Monica
  Palmirani}, {and} \bibinfo{person}{Fabio Vitali}.}
  \bibinfo{year}{2020}\natexlab{}.
\newblock \showarticletitle{Legal Knowledge Extraction for Knowledge Graph
  Based Question-Answering}. In \bibinfo{booktitle}{\emph{Legal Knowledge and
  Information Systems - {JURIX} 2020: The Thirty-third Annual Conference, Brno,
  Czech Republic, December 9-11, 2020}} \emph{(\bibinfo{series}{Frontiers in
  Artificial Intelligence and Applications}, Vol.~\bibinfo{volume}{334})},
  \bibfield{editor}{\bibinfo{person}{Serena Villata}, \bibinfo{person}{Jakub
  Harasta}, {and} \bibinfo{person}{Petr Kremen}} (Eds.).
  \bibinfo{publisher}{{IOS} Press}, \bibinfo{pages}{143--153}.
\newblock
\urldef\tempurl%
\url{https://doi.org/10.3233/FAIA200858}
\showDOI{\tempurl}


\bibitem[Sovrano and Vitali(2022a)]%
        {sovrano2022explanatory}
\bibfield{author}{\bibinfo{person}{Francesco Sovrano} {and}
  \bibinfo{person}{Fabio Vitali}.} \bibinfo{year}{2022}\natexlab{a}.
\newblock \showarticletitle{Explanatory artificial intelligence (YAI):
  human-centered explanations of explainable AI and complex data}.
\newblock \bibinfo{journal}{\emph{Data Mining and Knowledge Discovery}}
  (\bibinfo{year}{2022}).
\newblock
\showISBNx{1573-756X}
\urldef\tempurl%
\url{https://doi.org/10.1007/s10618-022-00872-x}
\showDOI{\tempurl}


\bibitem[Sovrano and Vitali(2022b)]%
        {sovrano2022how}
\bibfield{author}{\bibinfo{person}{Francesco Sovrano} {and}
  \bibinfo{person}{Fabio Vitali}.} \bibinfo{year}{2022}\natexlab{b}.
\newblock \showarticletitle{How to Quantify the Degree of Explainability:
  Experiments and Practical Implications}. In \bibinfo{booktitle}{\emph{31th
  {IEEE} International Conference on Fuzzy Systems, {FUZZ-IEEE} 2022, Padova,
  July 18-23, 2022}}. \bibinfo{publisher}{{IEEE}}, \bibinfo{pages}{1--9}.
\newblock


\bibitem[Sovrano and Vitali(2023)]%
        {SOVRANO2023110866}
\bibfield{author}{\bibinfo{person}{Francesco Sovrano} {and}
  \bibinfo{person}{Fabio Vitali}.} \bibinfo{year}{2023}\natexlab{}.
\newblock \showarticletitle{An objective metric for Explainable AI: How and why
  to estimate the degree of explainability}.
\newblock \bibinfo{journal}{\emph{Knowledge-Based Systems}}
  \bibinfo{volume}{278} (\bibinfo{year}{2023}), \bibinfo{pages}{110866}.
\newblock
\showISSN{0950-7051}
\urldef\tempurl%
\url{https://doi.org/10.1016/j.knosys.2023.110866}
\showDOI{\tempurl}


\bibitem[Speith(2022)]%
        {speith2022review}
\bibfield{author}{\bibinfo{person}{Timo Speith}.}
  \bibinfo{year}{2022}\natexlab{}.
\newblock \showarticletitle{A review of taxonomies of explainable artificial
  intelligence (XAI) methods}. In \bibinfo{booktitle}{\emph{Proceedings of the
  2022 ACM Conference on Fairness, Accountability, and Transparency}}.
  \bibinfo{pages}{2239--2250}.
\newblock


\end{thebibliography}

%

\end{document}